Topological data analysis suggests human brain networks reconfiguration in the transition from a resting state to cognitive load.

## AUTHORS

I.M. Ernston[1,2], A.A. Onuchin[3,4], T.V. Adamovich[2]

## AFFILIATION

[1] Lomonosov Moscow State University, Moscow, Russia
[2] Psychological Institute of the Russian Academy of Education, Moscow, Russia
[3] Laboratory of Complex Networks, Center for Neurophysics and Neuromorphic Technologies, Russia
[4] Skolkovo Institute of Science and Technology, Skolkovo, Russia

## ABSTRACT

***BACKGROUND:*** The functional network of the brain continually adapts to changing environmental demands. The environmental changes closely connect with changes of active cognitive processes. In recent years, the network approach has emerged as a promising method for analyzing the neurophysiological mechanisms that underlie psychological functions.

***AIM:*** This research aimed to explore how the topological features of functional connectomes in the human brain are linked to different cognitive states. The focus was on understanding the dynamic changes in brain networks during working memory tasks, with the goal of identifying network characteristics inherent to working memory. Such insights could improve our understanding of cognitive processes, especially related to working memory.

***METHODS:*** The present study examines topological characteristics of functional brain networks in resting state and in cognitive load, provided by the execution of the Sternberg Item Recognition Paradigm (SIRP) based on electroencephalographic data. EEG traces from 67 healthy adults were processed to estimate functional connectivity by the means of coherence method. We propose that the topological properties of the functional networks in the human brain are distinct between cognitive load and resting state with higher integration in the networks during cognitive load.

***RESULTS:*** It was shown that topological features of functional connectomes depend on the current state of cognitive processing and change with task-induced cognitive load variation. The analysis also demonstrated that functional connectivity during working memory tasks showed a faster emergence of homology groups generators, supporting the idea of a relationship between the initial stages of working memory execution and an increase in faster network integration, with connector hubs playing a crucial role.

***CONCLUSION:*** The study found that the topological features of functional brain networks are influenced by the type of cognitive process engaged by individuals and adapt to task changes. During working memory tasks, the functional connectivity exhibited a faster emergence of homology group generators, indicating a connection between the working memory execution and enhanced network integration. These results suggest that cognitive states, particularly related to working memory, are associated with distinct topological properties of functional brain networks, highlighting the importance of network dynamics in cognitive processing.

Cognitive neuroscience, Functional Neuroimaging, Brain Electrical Activity Mapping, Connectome Mapping, Working Memory

## INTRODUCTION

Network neuroscience, an approach aimed at analysis of the characteristics of structural and functional networks associated with cognitive functions, is one of the most promising ways of understanding psychophysiological mechanisms of cognition. Tremendous complexity of brain neuronal structures, abundant with interconnections, renders relation of cognitive functions to activity of separate brain regions inefficient. It seems that interactions between brain loci is a better way of understanding the brain mechanisms of cognition. Lately, a network approach that depicts the brain as a network of interconnected regions has become a popular way of brain activity analysis.

### Integration and segregation in networks

The widely used measure in the functional networks studies is the level of global segregation and integration in the network. Segregated networks are characterized with a more distinguished structure of processing modules, while integrated networks' nodes are much more interconnected. Functional segregation in the network is characterized by higher modularity and clearly distinguishable clusters of nodes in which the number of intracluster connections significantly exceeds the number of intercluster ones. The integrated network is characterized by low modularity and higher level of interconnection between all nodes of the network (see Figure 1). The new computational methodology has shown global topological properties of functional brain networks to have some unique features, such as small-worldness, that implies a low path length and high clustering, and provides an optimal ratio of the efficiency of information processing and the costs of its transmission for brain networks[1]. On top of that, there is convincing evidence for brain neural networks to present such complementary to global small-world architecture characteristics of topological organization as high clustering and high global efficiency[2] and highly modular community structure[3], which indicates a high number of nodes with multiple connections – network hubs[4].

Increasing the level of integration within the brain functional networks is often associated with cognitive activity[5]. The level of integration in the network can directly predict the performance in cognitive tasks, including the ones for working memory (WM), such as N-back task.

### Intrinsic connectivity

By applying the methods of network neuroscience, a large amount of data on the basic network architecture of the human brain, especially the cerebral cortex, was collected. Neuroimaging studies suggest that brain activity topologically is organized by functional networks, which are persistent in the states of cognitive load, as well as in the resting state. These networks are commonly referred to as intrinsic connectivity networks (ICN)[6] in order to denote functional brain networks detected regardless of current cognitive load, and to separate them from resting state networks. There is evidence that the topology of these functional networks is close to the anatomical neural topology of the corresponding brain regions[7–9], and these networks are associated with certain cognitive functions (e.g. networks of visual perception, long-term memory, cognitive control or attention[10]), supporting global information processing and other aspects of cognition. Key features of ICN organization of the human brain include 7 mostly distinguished networks according to the study by Yea et al, where the more stable networks were distinguished in fMRI data[11]. These networks are roughly specified as visual, somatomotor, dorsal attention, ventral attention, limbic, frontoparietal and default mode networks. Though some of these networks are suggested to aggregate multiple topologically and task-specific subnetworks, research suggests these networks can be attributed with specific functions. Nevertheless, these findings support the idea of topologically localized network organization of brain neural activity.

According to modern conceptions of brain network organization, the key elements of global brain networks are the highly connected zones in the brain, which are responsible for the transfer of information between specialized ICNs. Such zones, or hubs, can be either local (provincial hubs), connecting nearby nodes to a functional local sub-network, or global (connector hubs), through which local subnets communicate with each other. Studies show that global hubs form the "rich club"[12], which includes almost 70% of the shortest paths in the neural networks of the brain, due to which it is the most important element ensuring the effective operation of the global network[13]. The connection of the features of rich club networks with cognitive functions lies in the fact that local hubs, having many strong connections within their subnets, ensure the transition of the network to easily accessible states[14], in which internalized knowledge and experience are available for processing in various means.

**Neural network basis of working memory**

Working memory (WM) is one of the crucial cognitive functions that makes a significant contribution to the individual's cognition. The processes of encoding, storage, and retrieval information from memory are essential for many cognitive functions, including speech, reasoning, perception and motor activity[15].

A major role in the execution of working memory is attributed to the prefrontal cortex (PFC), especially by its dorsolateral part (DLPFC). The dorsolateral prefrontal cortex seems to be involved in information storage, in particular, about spatial positioning, while various parts of the ventral and lateral prefrontal cortex take part in storing non-spatial information (for example, objects, faces, words, etc.). On the other hand, each of these areas may have different functions, while the dorsolateral prefrontal cortex is involved in manipulating information, and the ventrolateral cortex suggested to be involved in its retention[16].

According to recent studies, during the WM task, brain networks have some specific properties, in particular, an increase in integration between the frontal-parietal and frontal-temporal lobes, and an increase in reconfiguration in the frontal regions is positively associated with the performance of memorization[17]. Latest findings suggest that the execution of a WM tasks leads to the increase of segregation in functional brain networks in comparison to the networks in the resting-state. The significance of the role fronto-parietal functional networks play in the WM performance is validated by the considerable accuracy of prediction models, based on the topological characteristics of the functional connectivity in these regions[18,19]. On top of that, Finc et al. showed that training has an impact on network segregation, induced by WM tasks: after training participants tend to have more modularity in functional networks, while subjects' performance in WM tests increases as well[20]. After the training, the integration between task-positive systems (frontoparietal, salience, dorsal attention, cingulo-opercular) increased while a decrease in the integration of the listed ICNs with the default-mode network was observed.

Considering the above mentioned facts, we hypothesize that the topological characteristics of functional networks of the human brain differ in working memory tasks and in resting state. Furthermore, we assume functional connectivity in working memory load to demonstrate more integrated organization with a distinct rich club of highly-connected hubs.

## RESEARCH AIMS

The aim of this research was to investigate the relationship between the topological features of functional connectomes in the human brain and cognitive processing. We sought to understand how the organization of functional networks within the brain changes depending on the specific type of cognitive task performed by the participants. One particular focus was on studying the patterns of functional connectivity during working memory tasks, which are crucial for temporary storage and processing of information in the brain. By analyzing the dynamic changes in the functional connectomes during different cognitive tasks, we aimed to identify potential associations between specific network characteristics and cognitive performance. This investigation provides valuable insights into the underlying mechanisms of cognitive processes, especially in relation to working memory, and contributes to a deeper understanding of brain functioning and its relevance to cognitive abilities.

## MATERIALS AND METHODS

The study involved 67 people aged 18-34 years ($m = 21.7$, $SD = 3.36$), 20 females and 47 males, all right-handed, with no known injuries and neurological disorders.

**WM task procedure**

The experiment involved 10 minute recording of resting state brain activity at intervals of 2 minutes with closed eyes and with open eyes consecutively, a total of 6 and 4 minutes of recording, respectively. After that, the participants were offered a task on working memory – Sternberg Item Recognition Paradigm (SIRP)[21]. In this paradigm, participants are shown sets of characters, in the present study – a sequence of 6 digits (the Sample stimulus), and then, after a certain delay, one character (the Control stimulus) is presented, and the participants must determine whether this character was a part of the original set (see Figure 2). Each task comprises the following stages:

1. presentation of a fixation cross in the center of the screen (presented for 0.5 second);
2. presentation of the Sample stimulus (2 sec);
3. postponement with repeated presentation of the fixation cross (2 sec);
4. presentation of the Control stimulus – Target or Non-target (0.5 sec);
5. time for participant's response with repeated presentations of the fixation cross (1.5 sec).

In total, 129 stimuli were presented to each participant.

The PsychoPy version 2023.1.1 was used to program the experiment, present stimuli to the participants and record behavioral data.

**Neurophysiological data acquisition and processing**

Brain activity was recorded with the 64–channel EEG system BrainVision actiCHamp by Brain Products GmbH. The recording was performed in a monopolar mode. The proprietary mounting of electrodes by BrainVision based on the 10-10 system was used, with FCz as a reference electrode, AFz – grounding electrode. An electromyogram was recorded using an EOG electrode placed under the right eye in order to correct artifacts from oculomotor musculature. Frequency range of electrical signal registration was 0.1 Hz - 1000 Hz.

During the pre-processing, electroencephalographic data were manually processed to remove major artifacts. After that, the recording sampling frequency was changed from 1000 Hz to 250 Hz, the frequency range was limited to 0 Hz and 50 Hz, and the reference electrode was changed from FCz to a virtual averaged reference. This stage of pre-processing procedure was performed by means of Brain Vision Analyzer 1.0 by Brain Products GmbH.

At the final stage of preprocessing, oculomotor and other artifacts were removed by ICA and damaged epochs and channels were restored using the Autoreject library for Python[22].

The localization of bioelectric signal sources was carried out in order to more accurately determine the features of the distribution of neural electrical activity in the brain. The location of the sources was determined using the "fsaverage" head and brain MRI model based on the "Buckner40" model. The "oct6" scheme was used (4098 points per hemisphere; the distance between sources is 4.9 $mm$; the area for each source is 24 $mm$).

To calculate the forward operator using the boundary-element model (BEM), areas with different conductivity are divided into triangular geometric units. For electroencephalographic data, 3 layers were used: the intracranial space, the skull and the scalp. After that, the BEM layers are assigned a conductivity value: for the scalp and parts of the brain, the default value was 0.3 *S/m*; for the skull – 0.006 *S/m*.

The activity of the sources was calculated using the dSPM[23] method. The result of the algorithm is an assessment of the activity of individual sources in the hemispheres (4098 per hemisphere), which were reduced to 75 zones in each hemisphere using the PCA method, the value of the first component was implemented. The zones in the brain cortex were extracted according to the Destrieux anatomical atlas[24]. The source localization procedures were performed using the MNE-Python 1.3.1 package.

Pre-processed EEG traces were subjected to a connectivity estimation for each subject and each condition (SIRP, eyes-closed and eyes-open resting state), and adjacency matrices were constructed using coherence method [25]. We estimated connectivity for the whole interval for each of the conditions, including SIRP execution. The spotlight of the present study was centered on the evaluation functional connectomes differences linked to different cognitive states, not different levels of working memory performance. Thus, no SIRP epochs were dropped, including the ones, which were recorded during unsuccessful trial of working memory task execution. In our study, we focused on the alpha (8 – 13 Hz) and beta (13 – 30 Hz) frequency bands. EEG data was divided into 6-second epochs with a 0.5-window overlap to capture temporal dynamics. Coherence was calculated within each epoch and frequency band and averaged firstly withing every single SIRP stimulus. Thereafter coherence values were averaged across stimuli within the same frequency band, and adjacency matrices were constructed to represent functional connectivity.

At the present time, there is no established consensus on the optimal value of the bond strength threshold for constructing adjacency matrices, despite the fact that this procedure is an established part of the connectivity analysis process. In recent studies of functional connectivity, authors more often use fairly high values of the threshold of the strength of connections, up to 0.8 - 0.95 [26,27]. However, it is believed that choosing a high threshold value can lead to the loss of a significant share of information, since weak functional connections in brain networks can play an important role in the neural mechanisms of cognitive functions[28]. In this study, a less conservative threshold value of 0.7 was used. Functional connectivity was estimated by means of the MNE-Python 1.3.1 package.

**Topological data analysis**

Topology is often colloquially described as representing the overall structure of data. Topological features are useful for capturing global, multi-scale, and intrinsic properties of datasets, in addition to more localized and rigid geometric features. The usefulness of topological features has been acknowledged with the emergence of topological data analysis (TDA). Many researchers have attempted to utilize this information to gain a new perspective on their datasets. In recent years, an extension of TDA has emerged, which involves integrating topological methods to enhance traditional data analysis.

One fundamental assumption in the field of data analysis is that data possesses a shape, meaning it is sampled from an underlying low dimensional manifold, which is referred to as the "manifold hypothesis" [29]. Instead of solely relying on statistical descriptions, TDA seeks to explore the underlying manifold structure of data sets in an algebraic manner. This involves computing descriptors of data sets that remain stable even when subjected to perturbations, and these descriptors encode intrinsic multi-scale information about the shape of the data.

Data shape turns out as a significant property especially in the field of network science. There are numeros works on the topological structure of different biologically inspired data: from structural [30] and functional connectomes, to eye movements [31] and single cell activities. TDA techniques have gained popularity in processing EEG signals because they can aid researchers in discovering new properties of complex and extensive data by simplifying the analysis by implementation of a geometrical approach.

**Fundamental TDA definitions**

Point clouds are a type of data representation in which data elements are represented as an unordered set of points in a Euclidean space with $n$ dimensions, denoted as $E^n$. A point cloud is defined as a finite subset of $E^n$. This type of data can be obtained from many natural experiments and can even be extracted from two-dimensional time series by disregarding the order of elements. The overall topology of point clouds can offer valuable insights into the structure of the data.

The typical approach to transforming the data points in a cloud $\{x_i\} \subseteq E^n$ into a single, unified topological object is to use them as vertices in a combinatorial graph. To determine the edges in the graph, an $\epsilon$-sized window of proximity is defined, such that two points $x_i$ and $x_j$ are connected by an edge if their distance $\rho(x_i, x_j)$ is less than or equal to $\epsilon$. However, this graph has a two-dimensional structure and cannot adequately capture the high-dimensional properties of the original space from which the data points were sampled. To overcome this limitation, a mathematical object known as a clique complex can be constructed on any graph object using a specific method of simplicial complex creation. Each clique on $n$ vertices in the graph are interpreted as an $(n - 1)$-dimensional combinatorial simplex. TDA methods work directly with these discrete constructions, but their topological properties can be generalized to topological simplices, which are the topological realizations of such combinatorial simplices. There are different methods of clique complexes construction, and the most commonly used and useful ones are the Delaunay, Vietoris-Rips (VR), Cech (C), and Alpha (A) complexes. These indexes are defined as follows:

Cech
$$Cech_r(X) = \{\sigma \subseteq X \mid \bigcap_{x \in \sigma} B_r(x) \neq \varnothing\}$$

Vietoris-Rips
$$VR_r(X) = \{\sigma \subseteq X \mid diam(\sigma) \leq 2r\}$$

Delaunay
$$Del(X) = \{\sigma \subseteq X \mid \bigcap_{x \in \sigma} V_x \neq \varnothing\},$$
$$V_x = \{y \in R^d \mid \|y - x\| \leq \|y - z\|, \ \forall z \in X\}$$

Alpha
$$Alpha_r(X) = \{\sigma \subseteq X \mid \bigcap_{x \in \sigma} (B_r(x) \cap V_x) \neq \varnothing\}$$

One of the central and main methods of TDA is called *persistent homology* and considers the existence of an ordered pair $(X, f)$, where $X$ is a set of data points and $f$ is a filter function defined in the domain of interest $X$. The filter function induces a filtration of the set $X$, i.e. a sequence of subspaces

$$\varnothing = X_0 \subseteq X_1 \subseteq X_2 \subseteq \ldots \subseteq X_n = X$$

which are often the sublevel sets $X = f^{-1}(-\infty, \varepsilon]$ of this function for an arbitrary threshold parameter $\varepsilon \in R$. For example, assume that $R = \{VR_i\}_1^N$ is a sequence of $VR$ complexes associated with a point cloud (data) for an increasing sequence of parameter values $\{\varepsilon_i\}_1^N$.

$$VR_1 \rightarrow VR_2 \rightarrow \ldots \rightarrow VR_N$$

where $i$ – inclusion maps between these complexes.

Persistent homology provides a tool for examining homology not for a single complex $VR_i$ but for a whole sequence of homology groups in each dimension * and for all $i < j$

$$\iota: H^*(VR_i) \rightarrow H^*(VR_j)$$

The dimensions of homology vector spaces named Betti numbers $\beta_i = dim(H^*(X))$ play a role of the most common topological invariants in data analysis practice.

Persistent homology is a method that allows us to track and uncover the emergence and disappearance of topological features in various dimensions during the filtration process, where a threshold parameter $\epsilon$ changes from $-\infty$ to $+\infty$. Persistent topological features are those that persist over a long interval of the threshold parameter, hence the name "persistent". The persistence of a homology group element is measured as the difference between the values of the filter function $f$ at its death and birth moments, denoted by $d_i$ and $b_i$, respectively. In other words, it quantifies how long the homology group element has existed and how important it is for the overall topology of the space.

A concise method for summarizing information about the lifespan of elements in homology groups of a particular filtration is typically achieved by using a "persistence diagram". These diagrams are used for any given dimension $k$ in the filtration, and provide a compact representation of the data descriptors.

$$\{D^k_{f(X)}\}_{k \in \{0...K\}}$$

which are called a $k$-th dimensional persistence diagram (PD)

$$D_{f^k}(X):=\{(b_i, d_i)\}^k_{i \in I}$$

where $\{(b_i, d_i)\}^k_{i \in I}$ is the multiset of birth-and-death intervals of topological features in the dimension $k$. Analogous way to think about PD is a multiset of points on the extended Euclidean plane $R^2 \cup \{+\infty\}$ in the birth-and-death coordinates.

To compare different PDs between each other, there are different metrics, i.e. bottleneck and Wasserstein distances. Given two persistence diagrams $D$ and $D'$, their bottleneck distance is defined as,

$$W_\infty(D, D') := inf_{\eta:D \rightarrow D'} sup_{x \in D} ||x - \eta(x)||_{\infty},$$

where $\eta: D \rightarrow D'$ denotes a bijection between the point sets of $D$ and $D'$ and $||.||_\infty$ refers to $L_\infty$ distance between two points in $R^2$.

Wasserstein distance is a generalization of bottleneck metric and defined as,

$$W_p(D_1, D_2) := \ inf_{\eta:D_1 \to D_2} (\sum_{x \in D_1} ||x - \eta(x)||_\infty^p)^{1/p}$$

In this work we used TDA to analyze the topology structure of the individual functional connectomes. Each functional connectome was interpreted as a simplicial complex (Vietoris Rips complexes mostly used in this study) and persistence diagrams in 0, 1 and 2 dimensions were computed for each of them. Pairwise bottleneck distances between VR-complexes were computed and used to construct point clouds of diagrams (see Figure 3). As a final step, point clouds were compared using Representation Topology Divergence (RTD) , the topological measure of complex data representations, such as point clouds, which have their own topological and geometrical structures [32].

To vectorize PDs it is possible to use different functions, which represent diagrams in the vector form, for example, Betti curves. Let $D$ be a persistence diagram. Its Betti curve is the function $\beta_i: R \to N$, where $R$ is a set of real numbers and $N$ is a set of natural numbers respectively, whose value on $s \in R$ is the number, counted with multiplicity, of points $(b_i, d_i)$ in $D$ such that $b_i \leq s < d_i$. The name is inspired from the case when the persistence diagram comes from persistent homology. The $\beta_i(s)$ describes the $ith$ Betti number or the count the independent $i$-cycles in each graph after all cliques have been filled in or $i$-dimensional 'holes' (A 1-cycle bounds a 2D area, a 2-cycle bounds a 3D volume etc).

All topological computations were made using the python gudhi 3.8.0 package [43].

## RESULTS

Times of birth and death of 1-dimensional and 2-dimensional holes and connected components during the filtration process provide significant information about graph structure and its possible functional roles. It was shown that PDs correspond to two states of similar cognitive activity level: resting state with closed eyes and resting state with open eyes form topologically and geometrically equal point clouds in 2-dimensional space, with RTD distance between them equal to 9. As opposed, the distance between point clouds formed by PDs corresponding to different cognitive states, drastically increases (for 25%) and equal to 11 (see Figure 4).

In addition, the speed of 2-dimensional and 3-dimensional hole deaths and appearance is significantly higher for functional networks during the SIRP. The topological structure of connectomes during the SIRP becomes equal to the resting state networks only at the 20000th filtration step (see Figure 5). This supports the reconfiguration hypothesis that states that functional networks of cognitively loaded tasks solving reconfigure faster.

On the other hand, it was shown that there is a little difference between networks in resting state with open or closed eyes. The general topology structure is quite similar, i.e. the times of the birth and death of the topological features does not differ during these states.

It was shown that real-world biological networks have significantly distinguished clique topologies in comparison with the random networks [42]. If a correlation matrix is not random it can uncover the "geometric" structure of data and indicates that neurons encode geometrically organized stimuli. To test statistical significance of found topological properties, we have generated random distance matrices on the same number of vertices as the number of vertices in the functional connectomes and computed their Betti curves with similar dimensions (see Figure 5).

## DISCUSSION

This study utilized a graph theory-driven approach to examine the complex causality patterns derived from EEG recordings. The aim was to identify distinct topological properties of the neural networks associated with processing of information in working memory, as well as topological features of the resting-state networks, captured in close vicinity of the moment of the execution of the cognitive task. This was elicited during visual SIRT performed by healthy middle-aged adults.

It was shown that topological features, such as time of birth 0th, 1st and 2nd homology groups generators, i.e. network connected components, 1-dimensional and 2-dimensional holes, differ significantly depending on the current cognitive state. Furthermore, analysis has shown functional connectivity in working memory tasks to demonstrate a higher speed of homology groups generators appearance, providing evidence in favor of links between early phases of working memory execution and increased global integration in functional networks. Moreover, since connector hubs are the nodes of high participation in global network interconnections, we hypothesize that they contribute the most to higher speed of birth and death of homology groups. Thus, described topological properties can be linked with hub-based network configuration in cognitive load.

These findings suggest that topological data analysis, performed on the EEG-derived functional connectivity, can represent the complexity of functional networks underlying the cognitive functions, including the working memory, highlighting the peculiar properties of topological features of brain networks in resting state as well as selectivity to of dynamics occurring during processing of memory items.

**Networks in resting state and in cognitive load.** Our data suggests that functional networks of the human brain demonstrate significantly distinct topology depending on the current level of the cognitive load. Resting state networks were constructed from data acquired when participants had their eyes closed or open. It is known that whether an individual has an ability to visually inspect surrounding space while not experiencing any particular cognitive load, brain activity changes and adapts in response to the need to support neural processes of visual perception. These changes express some whole-brain features, such as suppression of alpha-band EEG activity, linked to preparatory visual attention [33]. From the complex network analysis perspective resting state brain activity with open or closed eyes suggests differences in properties connected to specific cognitive abilities networks. According to previous research, the coordinated activity between the cingulo-opercular (COn) and right-frontoparietal (RFPn) networks is associated with visual processing, resulting in increased integration in visual perception [34].

On the other hand, topological analysis, performed in present study, has shown little difference between networks in resting state with open or closed eyes. We speculate that observed insensitivity of homology groups generators death reflects stronger linkage of given network characteristics to high-level information processing, which takes place during dedicated problem solving, but not the background perception and processing.

Such results correlate with the predictions of the Global Workspace theory [35] according to which, when accessing the global workspace, the flow of information becomes an object of conscious processing, it is available for conscious report and flexible behavior control. At the same time, getting into the workspace enhances this flow relative to others, which are additionally inhibited. In many cognitive models, the concept of workspace is associated with arbitrary attention and working memory, so that the limits on the capacity of the workspace correspond to the limits usually set for focal attention or working memory [36]. A dynamic model of workspace formation was proposed, according to which the community structure of locally synchronized modular subsystems for unconscious processing can be functionally rearranged by the launch of a globally synchronized system representing a consciously processed stimulus [37]. Such dynamic transitions from modular to global synchronization have been demonstrated in computational models, including using data on the structure of anatomical networks of humans or model animals [38]. Thus, observed in our study, higher rapidity of birth 0th and 1st homology groups generators in working memory task execution may reflect the above-mentioned process of conscious processing of information in the workspace, which takes place alongside with the suppression of other data flows.

**Features of network topology in working memory tasks**

The human brain is a complex organ that is capable of reorganizing and adapting itself in response to changes in the environment. Collected evidence suggests ICNs as the functional basis of cognitive functions, with specific global states related to cognitive performance [39]. Our results support this hypothesis, showing faster local networks integration during working tasks processing in contrast with lesser early functional network integration in resting states. The greater number of new connected components in functional connectome on early stages of working memory task execution can be interpreted as a process of a faster establishment of high-degree hubs in functional networks. Thus, networks reconfigure faster to a topology with a more expressed highly connected core of the "rich club". Based on the fact that the distribution of the distribution of quantity of connected components over time is asymmetrical with shift to the left, we may suggest a greater number of new nodes are included in the "rich club", moving from the status of provincial hubs to global hubs already in the early stages of information processing in working memory. Such features of network reconfiguration can indicate processes in the brain, during which certain ICNs are included in the global functional network. The connection of changes in topology with the phase of information processing in working memory suggests that the described process is specific depending on the cognitive function performed, which is consistent with the concept of ICNs specific to various cognitive functions.

These findings correspond with recent network neuroscience studies, suggesting that a more globally integrated network with less specialized segregation may be effective in sustaining working memory [40]. The content of working memory is defined by the interaction between selective perceptual information processing (such as visual or auditory information) operated via selective attention, and long-term memory (LTM) representations that are in a particular state of "accessibility" and require persistent activity of specialized networks controlled by attentional processes [41]. Therefore, a whole-brain network with high global information transfer (integration) may better sustain an optimal interplay between locally specialized networks, as seen in the local organization of working memory subnetworks.

## Conclusions

This study used topological data analysis to analyze EEG recordings and identify topological properties of neural networks involved in working memory processing. SIRT was performed by healthy middle-aged adults. The research examined resting-state networks as well as connectivity in cognitive load to understand topological features, specific to each of the states and nature of network reconfiguration in transition between the states.

The study revealed a significant association between the topological characteristics of functional connectomes and the level of cognitive load undertaken by the subjects. Particularly during working memory tasks, the analysis indicated a quicker emergence of homology group generators, suggesting a connection between the execution working memory tasks and enhanced rapid integration of networks.

Overall, the study suggests that topological data analysis can represent the complexity of functional networks underlying cognitive functions, including working memory. The research highlights the unique properties of topological features of brain networks in the resting state, as well as selectivity to dynamics occurring during processing of memory items.

**TEMPLATE**: Original Article

### ADDITIONAL INFORMATION

**Funding source.** This study was not supported by any external sources of funding.

**Patients' consent.** Written consent was obtained from all the study participants before the study screening in according to the study protocol approved by the bioethics committee of the Lomonosov Moscow State University.

**Competing interests.** The authors declare that they have no competing interests.

**Author contribution:**
I.M. Ernston – neurophysiological data acquisition and processing, visualization, collection and analysis of literary sources, writing and editing; A.A. Onuchin – neurophysiological data processing, visualization, preparation and writing of the text of the article; T.V. Adamovich – curation, neurophysiological data processing, editing. All authors confirm that their authorship meets the international ICMJE criteria (all authors have made a significant contribution to the development of the concept, research and preparation of the article, read and approved the final version before publication).

**ИНФОРМАЦИЯ ОБ АВТОРАХ**

| | |
|---|---|
| **Автор, ответственный за переписку:** | |
| **Эрнстон Илья Максимович, стажер;** <br> адрес: 125009, Москва, ул. Моховая, д. 11, стр. 9. <br> ORCID: <br> https://orcid.org/0009-0001-8465-8373 <br> eLibrary SPIN: 3968-9518 <br> e-mail: ilya.ernston@gmail.com <br> телефон: 9629997393 | **Ernston Ilia Maksimovich, research intern;** <br> address: Mokhovaya ulitsa, 11S9, Moscow, Moscow City, 125009, Russia <br> ORCID: <br> https://orcid.org/0009-0001-8465-8373 <br> eLibrary SPIN: 3968-9518 <br> e-mail: ilya.ernston@gmail.com |
| **Соавторы (должны быть приведены в порядке их перечисления в списке авторов рукописи):** | |
| **Онучин Арсений Андреевич, младший научный сотрудник;** <br> ORCID: <br> https://orcid.org/0000-0002-7811-5831; <br> eLibrary SPIN: 4081-5605; <br> e-mail: onuchinaa@my.msu.ru | **Arsenii Andreevich Onuchin, researcher;** <br> ORCID: <br> https://orcid.org/0000-0002-7811-5831; <br> eLibrary SPIN: 4081-5605; <br> e-mail: onuchinaa@my.msu.ru |
| **Адамович Тимофей Валерьевич, младший научный сотрудник;** <br> ORCID: <br> https://orcid.org/0000-0003-1571-9192 <br> eLibrary SPIN: 3897-2897; <br> e-mail: tadamovich11@gmail.com | **Adamovich Timofey Valerievich; researcher;** <br> ORCID: <br> https://orcid.org/0000-0003-1571-9192 <br> eLibrary SPIN: 3897-2897; <br> e-mail: tadamovich11@gmail.com |

**TABLES**

**FIGURES**

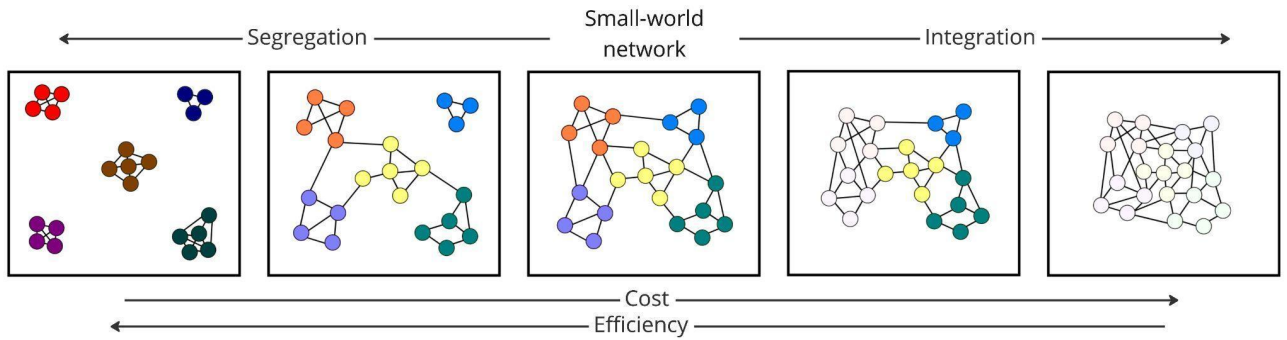

FIGURE 1 | Brain networks demonstrate a "small-world topology", providing a balance between a regular network (*the leftmost*), which promotes local efficiency in exchange for low costs, and a random network (*the rightmost*), which delivers global efficiency at high cost. As segregation increases (*right-to-left*), the network is divided into modules, the nodes of which are closely interconnected and poorly connected to the nodes in other modules. As integration increases (*left-to-right*), the number of connections between nodes increases, and individual modules merge into a single undifferentiated network. Rich club (*shown in yellow*), formed by hubs of high centrality, provides global information pathways in the network. Figure adapted from [6].

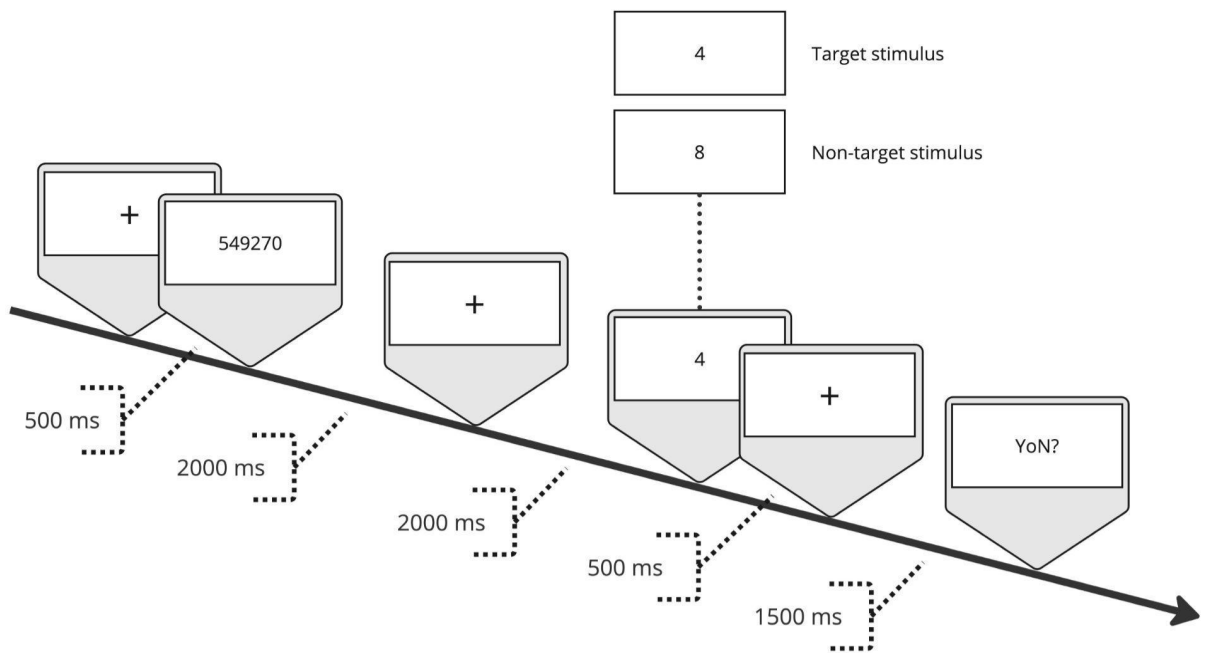

FIGURE 2 | Scheme of presentation of stimuli of the SIRP working memory task.

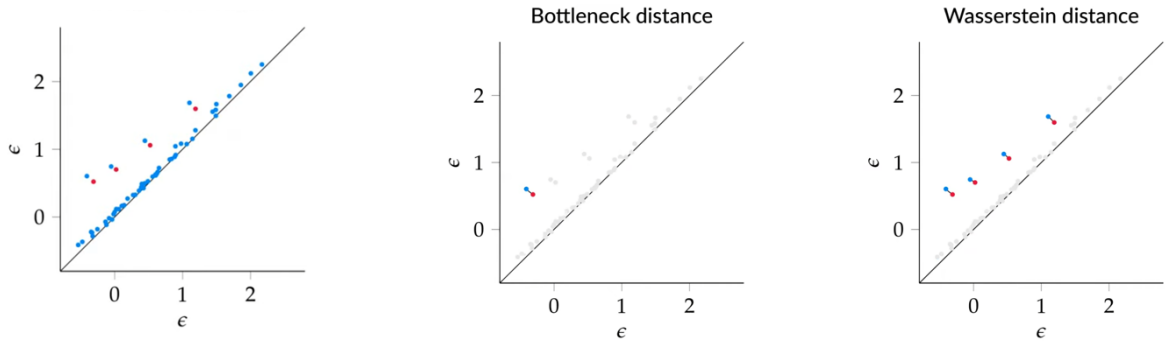

FIGURE 3 | Left picture represents the combination of two persistence diagrams: $D_1$ (blue) and $D_2$ (red), with the dots corresponding to moments of the homology vector space basis vectors deaths at specific $\epsilon$ filtration value moments. Two right pictures represent the difference between the information taken into the account in Bottleneck and Wasserstein distances between diagrams.

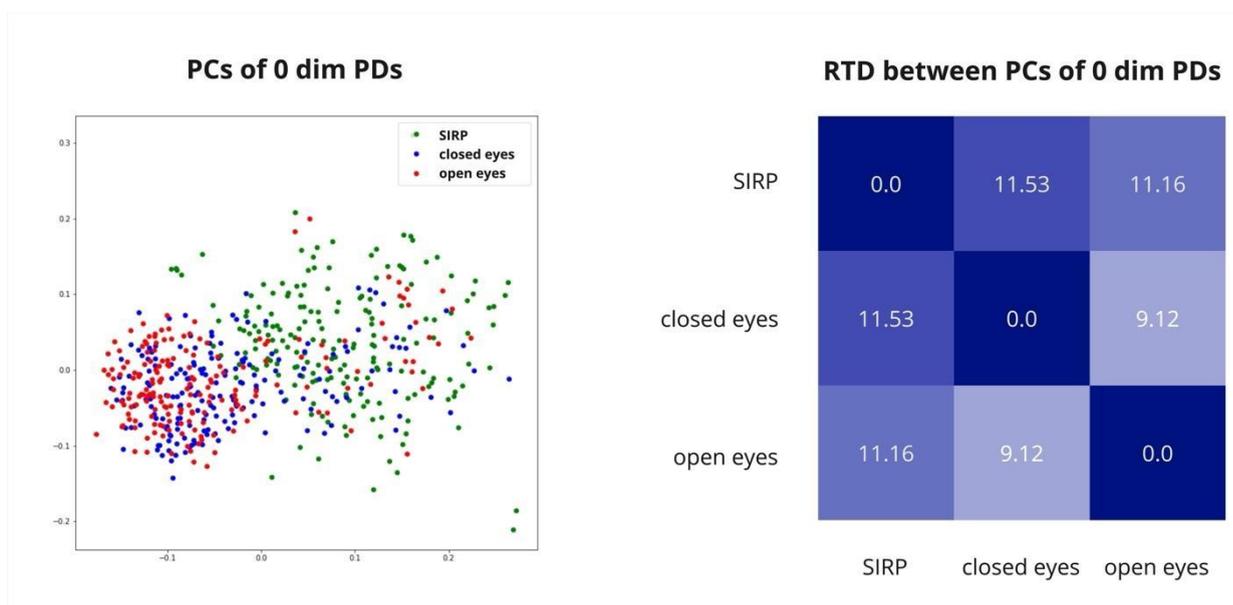

FIGURE 4 | Left picture shows the resulting point clouds for each of the three functional states (SIRP, closed eyes and open eyes) and clearly demonstrates the spatial discrepancy between persistence diagrams (points in the clouds), i.e. the cloud corresponding to the SIRP functional state is located far from resting state clouds (closed and open eyes functional states). Right picture shows the matrix of pairwise distances (Representation Topology Divergence metric[32]) between point clouds.

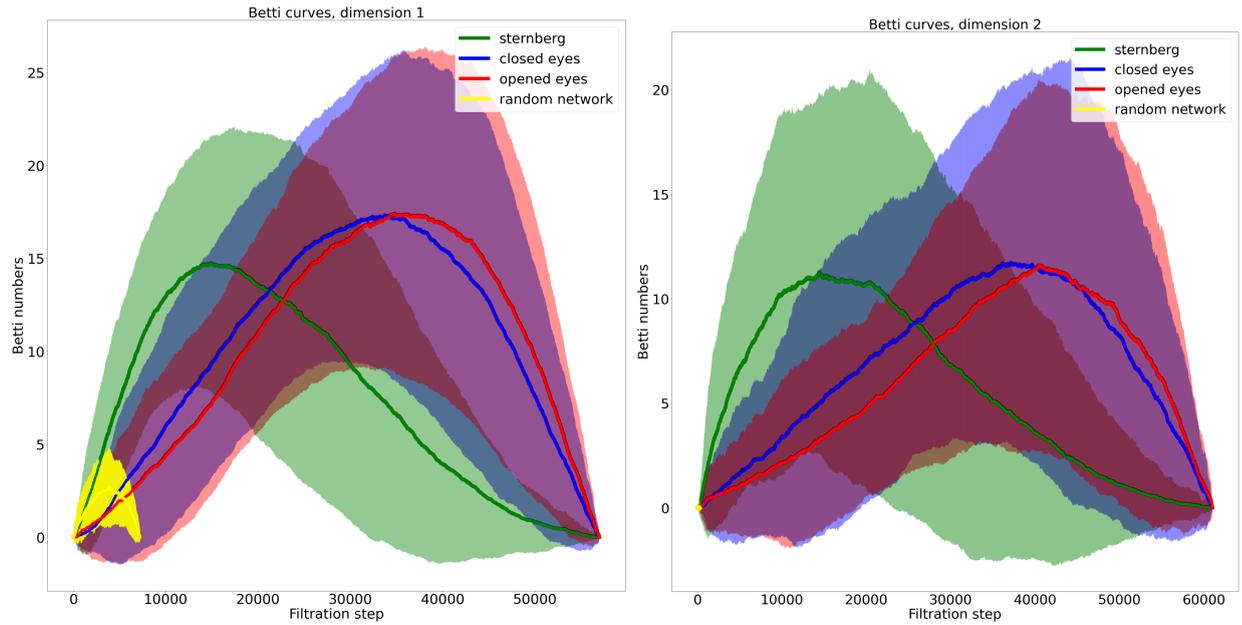

FIGURE 5 | Mean Betti curves for the dimensions 1 and 2 which are averaged across all participants. Y axis corresponds to the Betti numbers of the specific dimension: 1 or 2; X axis corresponds to the filtration steps for an increasing sequence of parameter values $\{\varepsilon_i\}_1^N$, where $N = 60000$ .